\documentclass[aps,onecolumn,superscriptaddress,nofootinbib]{revtex4-2}
\usepackage{graphicx}
\usepackage{dcolumn}
\usepackage{bm}
\usepackage[utf8]{inputenc}
\usepackage{float}
\usepackage{hyperref}
\usepackage{amsmath, amsthm, amsfonts,amssymb}
\usepackage[svgnames]{xcolor}
\usepackage{mathtools}
\usepackage{caption}
\usepackage{subcaption}
\usepackage{xcolor}
\usepackage{soul}
\usepackage{xcolor}
\usepackage{appendix}
\usepackage{orcidlink}
\usepackage{booktabs}
\usepackage{comment}

\definecolor{teal}{RGB}{0, 128, 128}
\hypersetup{ colorlinks=true, linkcolor=teal, citecolor=teal, urlcolor=teal}

\begin{document}

\title{Thermodynamics and quasinormal modes of the regular Dymnikova-Letelier black hole}

\author{L. C. N. Santos \orcidlink{0000-0002-6129-1820}}
\email{luis.santos@ufsc.br}
\affiliation{Departamento de Física, CFM - Universidade Federal de \\ Santa Catarina; C.P. 476, CEP 88.040-900, Florianópolis, SC, Brazil}

\author{L. G. Barbosa \orcidlink{0009-0007-3468-3718}}
\email{leonardo.barbosa@posgrad.ufsc.br}
\affiliation{Departamento de Física, CFM - Universidade Federal de \\ Santa Catarina; C.P. 476, CEP 88.040-900, Florianópolis, SC, Brazil}

\begin{abstract}
In this work, we investigate the thermodynamic properties and quasinormal modes of a regular Dymnikova-Letelier black hole. Starting from the Einstein field equations sourced by an effective anisotropic fluid, we analyze the resulting spacetime geometry and derive the associated thermodynamic quantities, including the Hawking temperature, heat capacity, and Gibbs free energy. The thermodynamic analysis reveals the existence of phase transitions characterized by divergences in the heat capacity, whose location depends sensitively on the string fluid parameter. 
We then study the dynamical response of the system under scalar perturbations by computing the quasinormal mode spectrum using the sixth-order WKB approximation. Our results show that, for all considered values of the parameters, the imaginary part of the quasinormal frequencies remains negative, while the real part stays positive, indicating the stability of the black hole under scalar perturbations. Furthermore, the presence of the string fluid leads to systematic shifts in both the oscillation frequencies and damping rates.
These results demonstrate that the string fluid significantly affects both the thermodynamic behavior and the dynamical stability of the Dymnikova-Letelier spacetime.
\end{abstract}

\maketitle

\section{Introduction}\label{Introduction}

Black holes provide a unique arena where geometry, matter content, and fundamental physical principles intertwine in a highly nontrivial way. Beyond their role as exact solutions of Einstein’s equations, black holes act as theoretical laboratories for probing the interplay between gravitation, thermodynamics, and field dynamics in strong-curvature regimes. In this context, regular black hole solutions sourced by effective matter fields have attracted considerable attention, as they evade curvature singularities while preserving key gravitational features such as horizons and thermodynamic behavior. Among these, the Dymnikova black hole \cite{dymnikova1992vacuum} stands out as a particularly appealing model, offering a smooth interpolation between a de Sitter core and an asymptotically Schwarzschild spacetime. When embedded in a string fluid or cloud environment, this geometry provides a natural framework to investigate how extended matter sources influence black hole thermodynamics and dynamical stability. It is important to emphasize that simply combining a string fluid source with the anisotropic fluid that supports the Dymnikova solution does not, by itself, guarantee a regular spacetime geometry. To achieve regularity, one can instead introduce a modified string fluid endowed with an variable equation-of-state parameter, which allows for a consistent interpolation between the matter content and the regular core structure \cite{santos2025revisiting}.

Regular black holes have been extensively investigated as physically motivated alternatives to singular solutions, providing spacetimes that are free from curvature singularities while preserving event horizons and consistent thermodynamic behavior \cite{Ditta:2025csz,Mustafa:2025cou,Anand:2025mlc,Waseem:2025bwb,Mannobova:2025uqf,Shahzad:2025msa,Vacaru:2025ngf,Calza:2025mrt,Mohamed:2025iqn,Song:2025qpo,Singh:2025ald,Alonso-Bardaji:2025qft,Kar:2025phe,Urmanov:2025nou,Mushtaq:2025ewk,Lin:2025zea,Lutfuoglu:2025hwh,Fernandes:2025eoc,Vertogradov:2025jxp,Capozziello:2025ycu,Ladghami:2025qjx,Huang:2025uhv,Khoo:2025qjc,Turakhonov:2025ojy,Coviello:2025pla,Barenboim:2025ckx,Ghaderi:2025xxw,Abu-Saleem:2025gfj,Harada:2025cwd,NunesdosSantos:2025alw,Casadio:2025pun,Javed:2025bpr,Konoplya:2025hgp,Borissova:2025msp,Narzilloev:2025nof,Khoshrangbaf:2025bwg,Chaudhary:2025lzi}. It is worth noting that solutions associated with black holes can be constructed so that the energy density remains regular at the origin, and may be employed in the description of compact objects surrounded by generic dark matter profiles \cite{figueiredo2023black}. In parallel, quasinormal modes (QNMs) have emerged as a powerful tool to probe the dynamical stability and observational signatures of black holes, encoding information about the underlying geometry and matter content through their characteristic oscillation spectra
Within the framework of general relativity, QNMs have been extensively investigated for a wide class of compact objects, establishing a robust theoretical foundation and enabling precise comparisons with observations \cite{Lan:2025brn,Ahmed:2025usm,Deng:2025hfn,Yin:2026npm,Liu:2025qos,Lutfuoglu:2025kqp,Lutfuoglu:2025pzi}. In parallel, the study of QNMs has emerged as a powerful probe of modified theories of gravity, where deviations from general relativity can leave distinctive imprints on the quasinormal spectrum, offering potential avenues to test gravity in the strong-field regime \cite{deOliveira:2025yeo,Sekhmani:2025zdc,Sekhmani:2025sml,Zahid:2025cfu,Hu:2025mvm,Lutfuoglu:2025blw}. For a comprehensive overview of analytic methods, theoretical foundations, and key results on black hole quasinormal modes, we refer the reader to the review article \cite{Bolokhov:2025rng}.

The study of black hole thermodynamics represents a cornerstone of modern gravitational physics, providing a powerful link between classical general relativity, quantum mechanics, and statistical mechanics. Since the seminal works of Bekenstein and Hawking established that black holes possess temperature and entropy \cite{bekenstein1973black, hawking1974black}, The analysis of thermal properties, such as Hawking temperature, entropy, and heat capacity, has become an indispensable tool for investigating the underlying microstructure and stability of black hole solutions. This thermodynamic framework allows for the classification of phases of gravitational configurations, where the sign of the heat capacity serves as a criterion for local stability, and its divergences typically indicate second-order phase transitions between stable and unstable branches.  This phase behavior has been extensively investigated in various environments, for example in black holes and black strings surrounded by quintessence \cite{Kiselev:2002dx,Barbosa:2025scy}, revealing rich structures dependent on the surrounding matter fields. Recent studies have examined thermodynamics and quasinormal modes of the D‑dimensional Dymnikova black hole \cite{Macedo:2024dqb}. Inspired by the same regularization mechanism, a regular string fluid was proposed \cite{santos2025revisiting}, which asymptotically reduces to the Schwarzschild black hole surrounded by a string cloud, forming the so-called regular Dymnikova–Letelier black hole. It is therefore a logical and timely step, offering a novel setting to investigate how the string‑fluid environment modifies both thermal stability and dynamical response. The Dymnikova--Letelier black hole provides a physically well-motivated framework by combining two relevant extensions of classical solutions: the regular Dymnikova core, which removes the central singularity, and the Letelier string-fluid, which models a distributed matter source with potential astrophysical relevance. In this sense, it represents a regular counterpart of black holes surrounded by a string fluid. This setup allows us to disentangle the effects of regularization (controlled by $r_0$) from those of the external matter sector (controlled by $\epsilon$). This suggests that the string-fluid sector may play a nontrivial role in shaping the ringdown signal, providing a useful framework for further investigation.

This paper is organized as follows. In Sec.~\ref{Field_equation_for_a_fluid_of_strings} we review the regular Dymnikova–Letelier black hole, obtained in Ref.~\cite{santos2025revisiting}. Section~\ref{Thermodynamics} studies the thermodynamic quantities of the Dymnikova solution with a string cloud: we compute the Hawking temperature, the heat capacity, and the Gibbs free energy, and we discuss the corresponding critical radii for representative values of $\epsilon$. In Sec.~\ref{Quasinormal_modes} we analyze massive scalar perturbations by calculating the quasinormal modes via the WKB method. Finally, in Sec.~\ref{Discussion_and_conclusions} we present a discussion and draw our conclusions.

\section{Field equation for a fluid of strings}\label{Field_equation_for_a_fluid_of_strings}
We consider the metric of a spacetime whose undetermined functions depend only on the radial coordinate, written in the form 
\begin{equation}
    ds^2 = -f(r) dt^2 + \frac{1}{f(r)} dr^2 + r^2 \left( d\theta^2 + \sin{\theta}^2 d\varphi^2 \right).
    \label{e1}
\end{equation}
The components of the energy-momentum tensor for a fluid of strings can be written as 
\begin{equation}
    T_{\: \: t}^{t} = T_{\: \: r}^{r}  \qquad \text{and} \qquad T_{\: \: \theta}^{\theta}=T_{\: \: \varphi}^{\varphi}= p.
    \label{e2}
\end{equation}
Considering the generalization proposed in \cite{santos2025revisiting}, the energy-momentum tensor for a fluid of strings is given by the expression
\begin{equation}
    T^{\mu}_{\;\;\nu}=\left[- \rho(r),- \rho(r),\frac{\rho(r)}{\alpha(r)},\frac{\rho(r)}{\alpha(r)}\right].
    \label{e3}
\end{equation}
By adopting the Einstein field equations, $G^{\mu}_{\:\:\:\nu} = 8\pi T^{\mu}_{\:\:\:\nu}$, together with the static and spherically symmetric metric introduced in (\ref{e1}) and the components of an anisotropic energy-momentum tensor (\ref{e3}), the resulting set of field equations can be written as follows:
\begin{align}
\frac{f'(r)}{r} + \frac{f(r)}{r^{2}} - \frac{1}{r^{2}} &= -8\pi \rho(r),\label{e4}\\
\frac{f''(r)}{2} + \frac{f'(r)}{r} &= 8\pi \frac{\rho(r)}{\alpha(r)}. 
\label{e5}
\end{align}
This equation can be easily integrated and the result is given by the expression 
\begin{equation}
f(r) = 1 + \frac{c_2}{r} + \frac{c_1}{r} \int e^{\left( \int \frac{-2}{\alpha(r)\, r} \, dr \right)} dr.
\label{e6}
\end{equation}
The equation above represents a formal solution of the field equations for an arbitrary function $\alpha(r)$. In what follows, we consider the solution associated with a string fluid, which can be obtained by adopting the following quantities
\begin{equation}
\alpha(r) = \frac{2 r_0^3 \left( \frac{\epsilon r_0^3 \exp\left(\frac{r^3}{r_0^3}\right)}{3} + \epsilon r^3 - \frac{\epsilon r_0^3}{3} + r_g r^2 \right)}{3 r^2 \left( \epsilon r^4 - \frac{4}{3} \epsilon r r_0^3 + r^3 r_g - \frac{2}{3} r_0^3 r_g \right)},
\label{e7}
\end{equation}
and 
\begin{equation}
    \rho(r) = \frac{ \left( \epsilon r_0^3 \exp\left(\frac{r^3}{r_0^3}\right) + 3\epsilon r^3 - \epsilon r_0^3 + 3 r_g r^2 \right)}{8\pi r_0^3 \exp\left(\frac{r^3}{r_0^3}\right) r^2}.
    \label{e8}
\end{equation}

The complete solution to the Einstein field equations, obtained by substituting the chosen equation of state and integrating, yields the following explicit form for the function $f(r)$:
\begin{align}
    f(r) = 1 - \left(\frac{r_g}{r} + \epsilon\right) \left(1-\exp\left(-\frac{r^3}{r_0^3}\right) \right),
    \label{e9}
\end{align}
with $c_1 = -1/r_0^3$, $c_2 = -r_g$ and $r_g=2M$. Equation (\ref{e9}) describes a regular black hole spacetime sourced by a fluid of strings. The parameter $r_0 > 0$ is a constant with dimensions of length that sets the regularization scale: for $r \gg r_0$, the exponential term becomes negligible and the solution reduces to the Schwarzschild geometry with a cloud of strings, $f(r) \approx 1 - 2M/r - \epsilon$. Near the origin, expanding the exponential reveals a de Sitter core $f(r) \approx 1 - (r_g/r_0^3) r^2$, ensuring that all curvature invariants remain finite at $r = 0$.

The energy density obtained from Einstein's equations can be manipulated to show that its limit at the origin is
\begin{align}
    \lim_{r \to 0} \rho(r) = \frac{3r_g}{8\pi r_0^3},
    \label{rhod0}
\end{align}
a finite value controlled by $r_0$. Smaller values of $r_0$ yield a denser core, while larger $r_0$ produce a more diluted configuration. In the limit $r_0 \to 0$, the central density diverges and we recover the singular Schwarzschild solution with a cloud of strings. Thus, $r_0$ acts as a regularity parameter that smooths out the classical singularity over a region of characteristic size $r_0$, establishing a de Sitter core near the origin while recovering the expected asymptotic behavior at large distances. The finiteness of both the energy density and the curvature invariants at $r = 0$ confirms that the spacetime described by Eq. (\ref{e9}) represents a genuine regular black hole solution \cite{santos2025revisiting}. Therefore, the main properties of the solution can be summarized as follows:
\begin{itemize}
  \item This geometry describes a \emph{regular string fluid} spacetime; however, in the asymptotic regime $r \gg r_0$, the solution becomes equivalent to that of a black hole surrounded by a \textbf{cloud of strings}.

  \item When $\epsilon = 0$, the solution reduces to the original Dymnikova solution.

  \item For $\epsilon = 0$ and in the limit $r \gg r_0$, the spacetime asymptotically approaches the Schwarzschild solution.

  \item The solution features a de Sitter core in the vicinity of $r = 0$.
\end{itemize}

In the following, we analyze in detail several physical aspects associated with this spacetime. Our investigation focuses on two complementary perspectives: the thermodynamic properties and the dynamical response under scalar perturbations. Both aspects are governed by the same parameters $\epsilon$ and $r_0$, allowing us to explore how the fluid of strings affects the black hole's equilibrium structure and its perturbative behavior. The analysis of Hawking temperature and heat capacity reveals phase transitions that signal changes in local thermodynamic stability, while the quasinormal mode spectrum provides a dynamical probe of the system's stability through oscillation frequencies and damping rates. Together, these complementary analyses offer a unified picture of how the string fluid shapes the physical properties of the regular Dymnikova-Letelier black hole.

\section{Thermodynamics}\label{Thermodynamics}
Now, we calculate the thermodynamic quantities associated with the regular Dymnikova-Letelier black hole described by (\ref{e9}). The parameter $M$ coincides asymptotically with the ADM mass, as can be verified from the behavior $f(r) \approx 1 - 2M/r - \epsilon$ for $r \gg r_0$. Assuming the existence of an event horizon $r_h$, the condition $f(r_h)=0$ allows us to express this parameter in terms of the horizon radius, resulting in
\begin{equation}
 M=\left(\frac{1}{\mathcal{D}}-\epsilon\right)\frac{r_{h}}{2},
   \label{e10}
\end{equation}
where, for convenience, we have introduced the auxiliary function
\begin{equation}
\mathcal{D} \equiv 1 - e^{-r_h^3/r_0^3},
\end{equation}
which encodes the exponential suppression arising from the regularization scale $r_0$. Given a black hole, we can associate a temperature with it, which we call the Hawking temperature, which in turn can be written as $T=\kappa/2\pi$ where $\kappa$ is the surface gravity given explicitly by
\begin{equation}
\kappa=\frac{1}{2}\left.\frac{df\left(r\right)}{dr}\right|_{r=r_{h}}.
\label{e11}
\end{equation}

By substituting the mass of the black hole, we can write the Hawking temperature as a function of the event horizon radius, the string cloud parameter, and the integration constant $r_0$. 
\begin{equation}
T=\frac{1}{4\pi r_{h}}\left(1-\epsilon\mathcal{D}-\frac{3}{\mathcal{D}}\frac{r_{h}^{3}}{r_{0}^{3}}e^{-\frac{r_{h}^{3}}{r_{0}^{3}}}\right).
 \label{e12}
\end{equation}

The equation (\ref{e12}) has the usual scaling $1/(4\pi r_h)$, the term in brackets combines the corrections of the Dymnikova core and the string cloud $\epsilon$, exponentially suppressed by the scale $r_0$ and which reduce to the asymptotic form $T=(1-\epsilon)/(4\pi r_{h})$.  
The local thermodynamic stability is governed by the heat capacity
\begin{equation}
C=\frac{dM}{dT},
\label{e13}
\end{equation}
The sign of $C$ serves as a direct stability criterion: $C>0$ corresponds to local stability, and $C<0$ to local instability, with the divergences $C \to \pm\infty$ marking second-order phase transitions that separate these stable and unstable branches.

Substituting the model expressions for $M$ and $T$ and simplifying, we obtain the following closed-form equation for the heat capacity
\begin{widetext}
\begin{equation}
  C=-\frac{2\pi r_{h}^{2}\left(1-\epsilon\mathcal{D}-\frac{3}{\mathcal{D}}\frac{r_{h}^{3}}{r_{0}^{3}}\mathrm{e}^{-\frac{r_{h}^{3}}{r_{0}^{3}}}\right)}{\mathcal{D}-\epsilon\mathcal{D}\left(\mathcal{D}-\frac{3r_{h}^{3}}{r_{0}^{3}}\mathrm{e}^{-\frac{r_{h}^{3}}{r_{0}^{3}}}\right)+\frac{6r_{h}^{3}}{r_{0}^{3}}\mathrm{e}^{-\frac{r_{h}^{3}}{r_{0}^{3}}}\left(1-\frac{3r_{h}^{3}}{2r_{0}^{3}}-\frac{3}{2\mathcal{D}}\frac{r_{h}^{3}}{r_{0}^{3}}\mathrm{e}^{-\frac{r_{h}^{3}}{r_{0}^{3}}}\right)}. 
\label{eq:heatcap}
\end{equation}
\end{widetext}

This expression explicitly defines how the string parameter \(\epsilon\) and the Dymnikova regularization scale $r_0$ govern the heat capacity. The exponential terms introduce a characteristic scale defined by $r_0$, while $\epsilon$ linearly modulates the intensity of the string cloud correction. 

In the large‑radius limit $r_{h} \gg r_{0}$, exponential corrections become negligible and the heat capacity reduces to the Schwarzschild‑like form $C = -2\pi r_{h}^{2}$, which is always negative, reflecting the familiar thermodynamic instability in that regime.

To characterize the nature of these phase transitions, we consider the Gibbs free energy~\cite{Errehymy:2025djk}, $G = M - T S$, where the entropy is given by the Bekenstein--Hawking area law $S = \pi r_h^2$. Using Eqs.~(\ref{e10}) and (\ref{e12}), we obtain the explicit expression
\begin{align}
   G=\frac{r_{h}}{4}\left(-1+\frac{2}{\mathcal{D}}-\epsilon\left(1+e^{-\frac{r_{h}^{3}}{r_{0}^{3}}}\right)+\frac{3}{\mathcal{D}}\frac{r_{h}^{3}}{r_{0}^{3}}e^{-\frac{r_{h}^{3}}{r_{0}^{3}}}\right). \label{eq:gibbs} 
\end{align} 

The functions $M(r_h)$, $T(r_h)$, and $S(r_h)$ are continuous for $r_h > 0$, and therefore $G$ is also continuous. Furthermore, $S = -dG/dT$ is continuous, as $S$ varies smoothly with $r_h$ and $T$ is a continuous function of $r_h$. The heat capacity $C = -T\,d^2G/dT^2$ exhibits divergences at points where $dT/dr_h = 0$, which delimit the regions of local stability ($C > 0$) and instability ($C < 0$).

This behavior exhibits similarities with that of a second-order phase transition: the function $G$ and its first derivative remain continuous, while the second derivative diverges. This type of criterion has been widely adopted in the context of black hole thermodynamics since the work of Davies~\cite{Davies:1977bgr}. In this sense, the divergences observed in $C$ can be interpreted as indicating second-order phase transitions, whose location is controlled by the parameter $\epsilon$, thus evidencing the influence of the string fluid on the thermodynamic structure of the system.

\begin{figure}[h!]
\centering \includegraphics[width=0.7\linewidth]{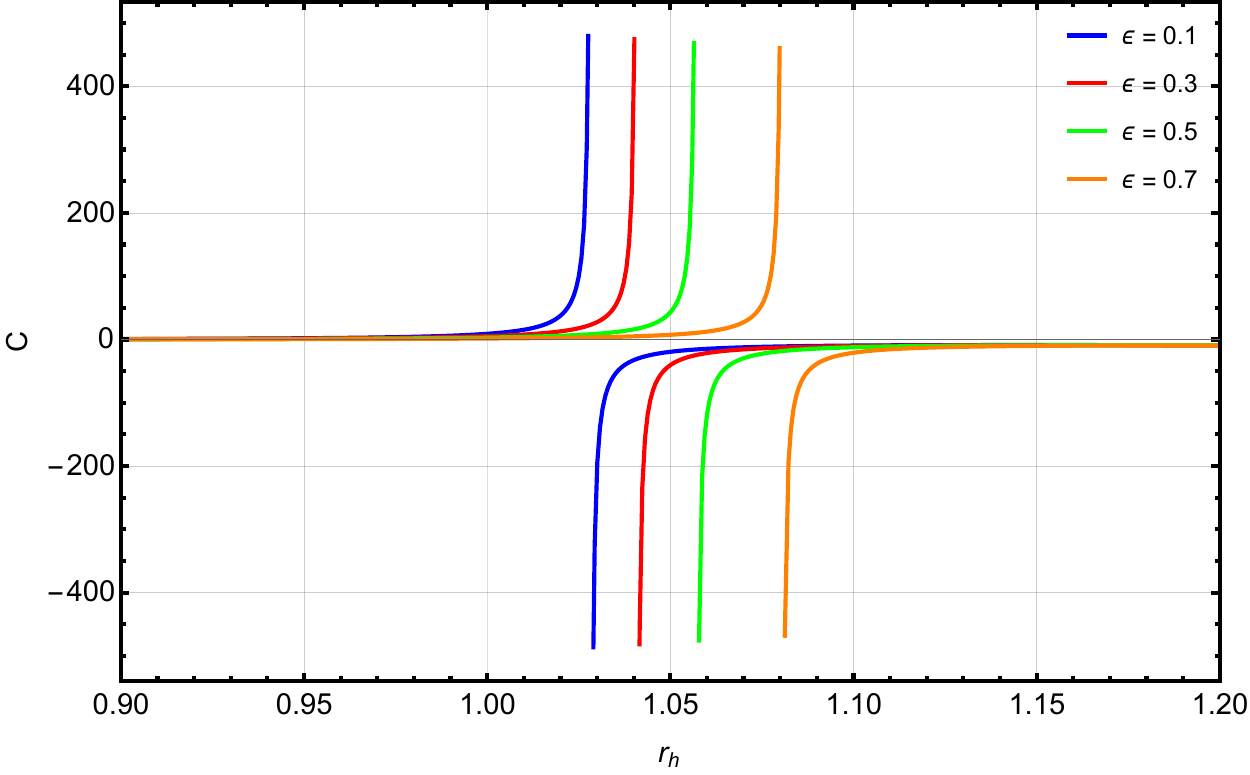}
\caption{Heat capacity \(C\) as a function of horizon radius \(r_{h}\) for \(r_{0}=0.5\). Curves correspond to \(\epsilon=0.1\) (blue), \(\epsilon=0.3\) (red), \(\epsilon=0.5\) (green) and \(\epsilon=0.7\) (orange). The plot was generated for \(r_{h}\in[0.1,2.3]\) and the displayed detail focuses on \(r_{h}\in[0.9,1.2]\).}
\label{fig:heat_capacity}
\end{figure}

As we can see in Fig. \ref{fig:heat_capacity}, the heat capacity exhibits divergences that separate intervals with positive $C$ from intervals with negative $C$. For each value of $\epsilon$ a pole appears near $r_{h}\sim1.0$ and systematically shifts to larger $r_{h}$ as $\epsilon$ increases, indicating that the threshold between local thermodynamic stability ($C>0$) and instability ($C<0$) depends on the string-cloud parameter. Away from these poles, and in particular for $r_{h}\gg r_{0}$, the exponential corrections become negligible and $C$ approaches the asymptotic form, recovering the standard unstable behaviour of large Schwarzschild-like black holes. Precise transition radii for given $\epsilon$ must be obtained by solving the denominator of the analytic expression for $C$.

\section{Quasinormal modes}\label{Quasinormal_modes}
Quasinormal modes (QNMs) are fundamental for characterizing how black holes respond dynamically when subjected to external disturbances. They correspond to damped oscillatory solutions with complex frequencies, where the real component sets the oscillation frequency and the imaginary component accounts for decay caused by energy dissipation. In this subsection, we analyze scalar QNMs in the regular fluid of string spacetime applying the WKB method to determine their spectra through an automated implementation \cite{Konoplya:2019hlu}.

Let us consider a scalar field $\Psi$ with mass $\mu$ propagating in a spacetime described by the metric $g_{\mu\nu}$. The Klein--Gordon equation for this system is given by
\begin{equation}
    (\nabla_{\nu}\nabla^{\nu} - \mu^2)\Psi = 0,
    \label{41}
\end{equation}
with $\nabla_{\nu}$ being the covariant derivative.
With some algebra and by using the properties of the covariant derivative, this equation can be cast into a form that facilitates the analysis of its solutions in curved geometries:
\begin{equation}
\frac{1}{\sqrt{-g}} \, \partial_\nu \left( g^{\mu\nu} \sqrt{-g} \, \partial_\mu \Psi \right) - \mu^2 \Psi = 0.
\label{42}
\end{equation}
We consider a massive scalar field $\Psi$ propagating in a static and spherically symmetric spacetime described by the line element (\ref{e1}). For this case, one has $\sqrt{-g} = r^2 \sin\theta$. Substituting this into the Klein--Gordon equation and expanding explicitly, we obtain
\begin{widetext}
    \begin{equation}
 -\frac{1}{f(r)} \frac{\partial^2 \Psi}{\partial t^2}
+ \frac{1}{r^2} \frac{\partial}{\partial r}\!\left(r^2 f(r) \frac{\partial \Psi}{\partial r}\right) 
+ \frac{1}{r^2 \sin\theta}
\frac{\partial}{\partial \theta}\!\left(\sin\theta \frac{\partial \Psi}{\partial \theta}\right)
+ \frac{1}{r^2 \sin^2\theta}\frac{\partial^2 \Psi}{\partial \phi^2}
- \mu^2 \Psi = 0 .
\end{equation}
\end{widetext}

Exploiting the spherical symmetry of the background, we perform a separation of variables of the form
\begin{equation}
\Psi(t,r,\theta,\phi) =
e^{-i\omega t}
R(r)
Y_{\nu m}(\theta,\phi),
\end{equation}
where $Y_{\ell m}$ are the spherical harmonics. Substituting this ansatz into the Klein–Gordon equation and considering the tortoise coordinate $r_*$ defined by
\begin{equation}
\frac{dr_*}{dr} = \frac{1}{f(r)},
\end{equation}
and redefining the radial function as
\begin{equation}
R(r) = \frac{u(r)}{r},
\end{equation}
the radial equation can be cast into a Schrödinger-like form:
\begin{equation}
\frac{d^2 u}{dr_*^2}
+
\left[
\omega^2 - V_\nu(r)
\right] u = 0.
\end{equation}

The effective potential is given by
\begin{equation}
V_\nu(r)
=
f(r)
\left[
\frac{\nu(\nu+1)}{r^2}
+
\frac{f'(r)}{r}
+
\mu^2
\right]
\label{potential}.
\end{equation}

This potential encodes the combined effects of spacetime curvature, angular momentum, and the scalar field mass, and plays a central role in the analysis of wave propagation and quasinormal modes in spherically symmetric spacetimes. We now turn to the method that will be employed to obtain the quasinormal modes. In this work, we use the WKB approximation, which has proven a semi-analytical technique for studying wave propagation in black hole spacetimes. The method is based on treating the radial scalar equation as a Schrödinger-like problem and approximating the solution near the peak of the effective potential. It allows one to extract the complex frequencies associated with damped oscillations, providing accurate results especially for overtone numbers.

We adopt the sixth-order WKB approximation \cite{Konoplya:2003ii}, in which the quasinormal frequencies are obtained from the effective potential (see Eq.~\ref{potential}) through the relation
\begin{equation}
    \frac{w^2 - V_0}{\sqrt{-2V_0^{''}}} - \Lambda_2 - \Lambda_3 - \Lambda_4 - \Lambda_5 - \Lambda_6 = n + \frac{1}{2},
\end{equation}
where $V_0$ denotes the maximum of the effective potential and $V_0^{''}$ its second derivative evaluated at that point. The explicit expressions for the correction terms $\Lambda_i$, with $2 \leq i \leq 6$, can be found in Refs.~\cite{Iyer:1986np,Konoplya:2003ii}. An automated implementation for computing the quasinormal modes using this method is available in \cite{Konoplya:2019hlu}.
In summary, the procedure consists of the following steps:
\begin{enumerate}
    
    \item Identify the maximum of the effective potential $V_{\text{eff}}(r)$ and compute its derivatives at this point, since the WKB approximation is constructed around this maximum.
    
    \item Apply the WKB formula at the desired order (sixth order), which relates the complex frequency $\omega$ to the value and derivatives of the potential at its peak.
    
    \item Solve the resulting algebraic equation for $\omega$, obtaining the quasinormal frequencies for a given overtone number $n$.
    
    \item Repeat the procedure for different values of the parameters of the model in order to analyze the behavior of the quasinormal spectrum.
\end{enumerate}

\begin{figure}[h!]
\centering \includegraphics[width=0.7\linewidth]{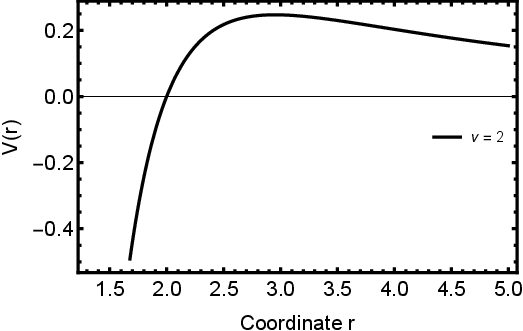}
\caption{The figure displays the effective potential $V(r)$ as a function of the radial coordinate $r$ for $M = 1$ and $r_0 = 0.3$. The potential presents a well-defined minimum near the horizon region, followed by a smooth increase and the formation of a potential barrier at larger values of $r$. 
}
\label{fig2}
\end{figure}
We begin the analysis of the results by examining the effective potential associated with the radial equation for a massless scalar field. For the set of parameters considered, it is clearly seen in Fig.~\ref{fig2} that the effective potential exhibits a pronounced peak around the region $r \approx 3.28$. This barrier structure is a typical feature in perturbative problems and plays a central role in determining the QNM spectrum. Physically, this peak acts as a potential barrier that partially reflects and partially transmits the scalar perturbations. As a consequence, perturbations trapped near this region give rise to damped oscillations that characterize the quasinormal mode spectrum. In many black hole spacetimes, the dominant contribution to the QNMs is governed by the properties of the effective potential near its maximum. Throughout the analysis, the overall shape of the potential remains qualitatively similar for different values of the parameters. This indicates that the qualitative features responsible for the QNM spectrum remain essentially unchanged across the considered parameter range, suggesting that the underlying structure of the spacetime is preserved.

\begin{figure}[h!]
\centering \includegraphics[width=0.7\linewidth]{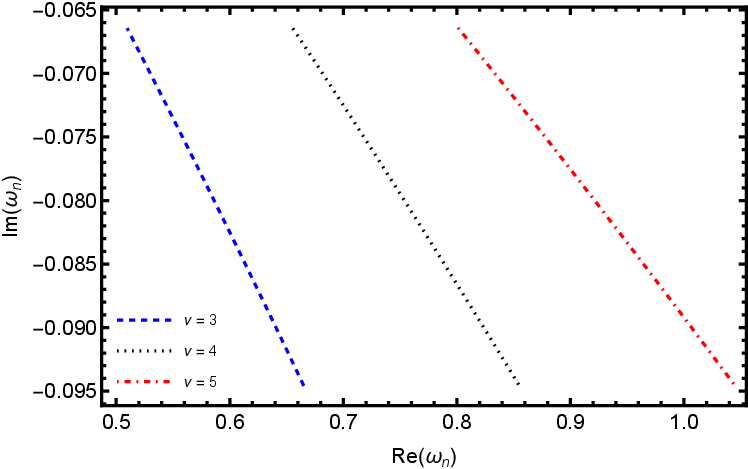}
\caption{QNM spectrum in the complex frequency plane for $n=0$, $M=1$, and $r_0=0.3$, computed for different values of the parameter $\epsilon$. The real and imaginary parts of the frequencies illustrate the dependence of the damping and oscillation rates on $\epsilon$. Distinct curves correspond to different choices of $\nu$, highlighting the systematic shift of the QNMs in the complex plane.}
\label{fig3}
\end{figure}

We now proceed to analyze the QNMs associated with the fundamental mode $n=0$. As illustrated in Fig.~\ref{fig3}, the imaginary part of the frequency remains negative, while the real part stays positive for values of the parameter $\epsilon$ ranging from $0.01$ to $0.017$. 
The positivity of the real part of the frequency corresponds to the oscillatory behavior of the perturbations, while the negative imaginary part indicates that these oscillations decay with time.
This behavior confirms the stability of the mode within the considered parameter interval. 

\begin{figure}[h!]
\centering \includegraphics[width=0.7\linewidth]{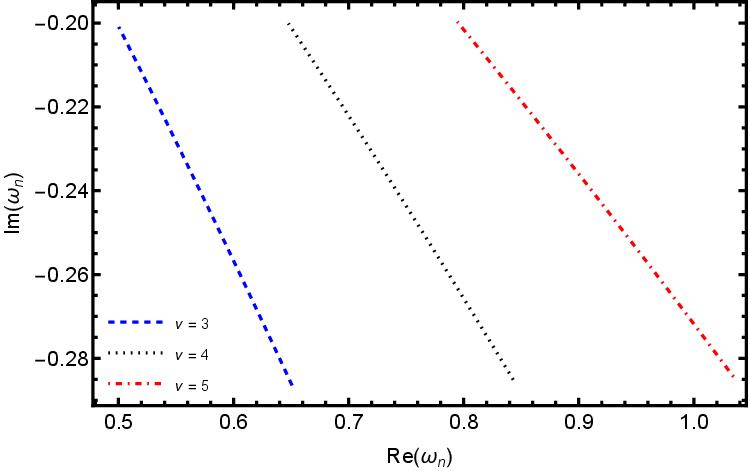}
\caption{Complex frequency spectrum for $n=1$ with $M=1$ and $r_0=0.3$, obtained for different values of the parameter $\epsilon$. The figure shows the displacement of the modes in the complex plane as $\epsilon$ varies, indicating changes in both oscillation frequency and decay rate. The distinct branches correspond to different parameter choices, revealing a systematic trend in the mode behavior.}

\label{fig4}
\end{figure}

Similarly, the QNMs associated with the overtone number $n=1$, shown in Fig.~\ref{fig4}, indicate that the imaginary part of the frequency remains negative while the real part stays positive for the considered values of $\epsilon$. In comparison with the fundamental mode, the magnitude of the imaginary part becomes larger, indicating that the corresponding perturbations decay more rapidly.

\begin{figure}[h!]
\centering \includegraphics[width=0.7\linewidth]{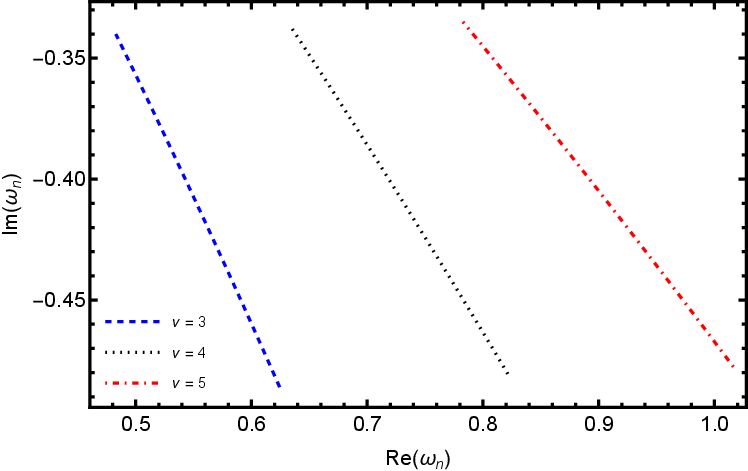}
\caption{Frequency spectrum in the complex plane for $n=2$, with $M=1$ and $r_0=0.3$, for several values of the parameter $\epsilon$. The figure illustrates the progressive shift of the modes as $\epsilon$ varies, showing a clear dependence of both the real and imaginary parts of the frequency on the chosen parameter. The separate curves indicate distinct branches of the spectrum.}

\label{fig5}
\end{figure}

Finally, Fig.~\ref{fig5} presents the QNMs corresponding to the overtone number $n=2$, which exhibit a behavior analogous to the previous cases, further indicating the stability of the system under scalar perturbations. As expected from the general structure of quasinormal spectra, higher overtones tend to have larger damping rates, reflecting the progressively shorter-lived nature of these modes.

Table \ref{tab:qnm_modes} compares the quasi-normal mode spectrum of the standard Schwarzschild black hole with that of the Dymnikova--Letelier black hole, keeping the Dymnikova core parameter $r_0$ fixed while varying the string-fluid parameter $\epsilon$. The results show that the parameter $\epsilon$ plays a crucial role in controlling deviations from the Schwarzschild geometry. For very small values of $\epsilon$ ($\epsilon = 0.0005$), the quasi-normal frequencies of the Dymnikova--Letelier solution remain close to the Schwarzschild case, especially for the $(n,l)=(1,1)$ mode. This indicates that when the string-fluid contribution is weak, the effective spacetime geometry is only slightly perturbed, and the oscillation and damping properties closely resemble those of the usual Schwarzschild black hole. However, when $\epsilon$ is increased to $\epsilon = 0.1800$, significant departures from Schwarzschild behavior emerge. In particular, the real part of the frequency increases dramatically, while the imaginary part decreases in magnitude. This means that the Dymnikova--Letelier black hole exhibits oscillations at much higher frequencies with slower damping rates. Such behavior suggests that the surrounding string-fluid component substantially modifies the effective potential barrier governing perturbations. From a physical perspective, this implies that the combined Dymnikova--Letelier geometry interpolates between two regimes: for small $\epsilon$, it behaves approximately like a Schwarzschild black hole, whereas for larger $\epsilon$, it develops a distinct ringdown signature characterized by enhanced oscillation frequencies and longer-lived modes. Such deviations may provide an observational channel to constrain exotic matter distributions surrounding regular black holes through gravitational-wave spectroscopy.
\begin{table}[ht]
\centering
\caption{Quasi-normal modes for Schwarzschild and Dymnikova--Letelier black holes for $r_0=0.3$.}
\begin{tabular}{ccccc}
\hline
$n$ & $l$ & $\epsilon$ & Schwarzschild \cite{Konoplya:2003ii} & Dymnikova--Letelier \\
\hline
0 & 0 & 0.0005 & $0.220928 - 0.201638\,i$ & $0.407080 - 0.094191\,i$ \\
1 & 1 & 0.0005 & $0.528942 - 0.613037\,i$ & $0.524607 - 0.617766\,i$ \\
1 & 1 & 0.1800 & $0.528942 - 0.613037\,i$ & $2.772240 - 0.127075\,i$ \\
2 & 1 & 0.1800 & $0.462028 - 1.084330\,i$ & $7.827710 - 0.167510\,i$ \\
\hline
\end{tabular}
\label{tab:qnm_modes}
\end{table}

\section{Discussion and conclusions}\label{Discussion_and_conclusions}

In this work, we have investigated the thermodynamic properties and QNM spectrum of a regular Dymnikova-Letelier black hole. By combining analytical methods with numerical analysis, we explored how the presence of the string fluid/cloud parameter $\epsilon$ and the regularization scale $r_0$ affect both the thermal behavior and the dynamical response of the system under scalar perturbations.

From the thermodynamic perspective, we derived explicit expressions for the Hawking temperature and heat capacity. The analysis suggests the existence of phase transitions characterized by divergences in the heat capacity, separating stable and unstable branches. The location of these critical points was shown to depend sensitively on the string fluid parameter, with larger values of $\epsilon$ shifting the transition radius to higher values. In the asymptotic regime, the thermodynamic behavior reduces to that of a Schwarzschild black hole, as expected, confirming the consistency of the model.

We then analyzed the QNM of a massless scalar field using the sixth-order WKB approximation. The effective potential exhibits a well-defined barrier, validating the applicability of the WKB approach. For all cases considered, including the fundamental mode and higher overtones, the imaginary part of the frequency remains negative, while the real part stays positive. This behavior confirms the dynamical stability of the system under scalar perturbations. Furthermore, the systematic shift of the QNM frequencies with increasing $\epsilon$ indicates that the string fluid significantly influences the oscillation and damping properties of the scalar perturbation.

Overall, our results show that the presence of a string cloud modifies both the thermodynamic structure and the quasinormal spectrum of the Dymnikova-Letelier black hole in a controlled manner. These findings reinforce the robustness of regular black hole solutions in extended matter configurations and highlight the relevance of QNM modes as probes of the stability under scalar perturbation. Future investigations may extend this analysis to gravitational perturbations, higher-dimensional configurations, or alternative modified gravity scenarios, providing further comprehension into the interplay between geometry, matter content, and black hole dynamics.

\section{Acknowledgements}\label{Acknowledgements}
LGB would like to thank CAPES (Process number: 88887.968290/2024-00) for the financial support. LCNS would like to thank Conselho Nacional de Desenvolvimento Científico e Tecnológico - Brazil (CNPq) for financial support under Research Project No. 443769/2024-9 and Research Fellowship No. 314815/2025-2.

\bibliographystyle{unsrturl}
\bibliography{sample}

\end{document}